\documentclass[preprint,nofootinbib,floats,superscriptaddress,tightenlines,amsfonts,amsmath,amssymb]{revtex4}

\usepackage{graphicx}
\usepackage{bm}
\usepackage{color}
\usepackage{hyperref}
\usepackage{multirow}
\usepackage{slashed}

\begin{document}
\baselineskip=17pt \parskip=5pt

\preprint{NCTS-PH/1908}
\hspace*{\fill}

\title{Searching for dark photons in hyperon decays}

\author{Jhih-Ying Su}
\affiliation{Department of Physics, National Taiwan University, Taipei 106, Taiwan}

\author{Jusak Tandean}
\affiliation{Department of Physics, National Taiwan University, Taipei 106, Taiwan}
\affiliation{Physics Division, National Center for Theoretical Sciences, Hsinchu 300, Taiwan
\bigskip}


\begin{abstract}

That massless dark photons could exist and have flavor-changing magnetic-dipole interactions
with down-type light quarks is an attractive possibility which may be realized in various
new-physics scenarios.
It is potentially testable not only in kaon processes but also via two-body hyperon decays
involving missing energy carried away by the massless dark photon.
We explore the latter within a~simplified model approach and take into account constraints
from the kaon sector.
We find that the branching fractions of some of these hyperon modes are allowed to be
as high as a few times $10^{-4}$.
Such numbers are likely to be within the sensitivity reaches of ongoing experiments like
BESIII and future ones at super charm-tau factories.

\end{abstract}

\maketitle

Models of new physics (NP) beyond the standard model (SM) may have a dark sector containing
an extra Abelian gauge group, U(1)$_D$, under which all SM fields are singlets.
This symmetry may be broken spontaneously or unbroken, causing the associated gauge boson,
the dark photon, to get mass or stay massless.
These possibilities are appealing for a number of reasons and have been much studied in
the past~\cite{Holdom:1985ag,Dobrescu:2004wz,Gabrielli:2016cut,Fabbrichesi:2017vma,
Fabbrichesi:2019bmo,Ackerman:2009mha,Zhang:2018fbm,He:2017zzr,Barger:2011mt,Chiang:2016cyf,
Batley:2015lha,Pospelov:2008zw,Essig:2013lka,Alexander:2016aln}.
They have motivated many dedicated hunts for dark photons in recent years~\cite{Essig:2013lka,
Alexander:2016aln,Ablikim:2017aab,Aaij:2017rft,Anastasi:2018azp,Ablikim:2018bhf,
CortinaGil:2019nuo,NA64:2019imj}, with negative outcomes so far.
The majority of these efforts have been based on the assumption that U(1)$_D$ is broken and
the accompanying massive dark photon,~$A'$, can interact directly with SM fermions through
a renormalizable operator $\epsilon eA_\mu'J_{\textsc{em}}^\mu$, with $eJ_{\textsc{em}}$ being
the electromagnetic current and the small parameter $\epsilon$ originating from kinetic mixing
between the dark and SM Abelian gauge fields~\cite{Pospelov:2008zw,Essig:2013lka,Alexander:2016aln}.
This entails that $A'$ could be produced in the scattering or decays of SM particles, including
the decays of unflavored mesons, and it could decay into electrically charged fermions or mesons
and perhaps also invisibly into other dark particles, allowing experiments to set limits on
$\epsilon$ over various ranges of the $A'$ mass~\cite{Essig:2013lka,Alexander:2016aln,
Ablikim:2017aab,Aaij:2017rft,Anastasi:2018azp,Ablikim:2018bhf,CortinaGil:2019nuo,NA64:2019imj}.

If instead U(1)$_D$ remains unbroken, its associated gauge boson will not acquire any mass.
In this case, there is always a linear combination of the dark and SM Abelian gauge fields which
does not have renormalizable couplings to SM members and which can be identified with the massless
dark photon~\cite{Holdom:1985ag,Dobrescu:2004wz}.
Thus, the latter, which we denote by $\overline\gamma$, has no direct interactions with SM
fermions, unlike the massive one.
As a consequence, $\overline\gamma$ can escape the restrictions implied by the aforementioned
quests for $A'$.
Nevertheless, since $\overline\gamma$  can still exert influence on the SM via higher-dimensional
operators generated by loop diagrams involving particles which are charged under U(1)$_D$ and
also coupled to SM fields \cite{Dobrescu:2004wz,Gabrielli:2016cut,Fabbrichesi:2017vma},
more efficient ways to look for it may be available and hence should be pursued.
Here we propose some of them, which apply to the fairly general situation described in
the following.

We concentrate on flavor-changing neutral current (FCNC) effects arising from the massless dark
photon having nonrenormalizable interactions with the $d$ and $s$ quarks.
This might bring to mind kaon processes which for the massive dark photon could serve as good
probes, especially the decays \,$K\to\pi A'$\, and \,$K^+\to\mu^+\nu A'$ \cite{Pospelov:2008zw,
Barger:2011mt,Chiang:2016cyf,Batley:2015lha}.
However, with $\overline\gamma$ being massless and not interacting directly with SM
fermions, \,$K\to\pi\overline\gamma$\, is forbidden by angular-momentum conservation and
gauge invariance and \,$K^+\to\mu^+\nu\overline\gamma$\, would have a very suppressed rate.
Rather, in this massless case, observing \,$K^+\to\pi^+\pi^0\overline\gamma$\, has been suggested
as an avenue to study the \,$ds\overline\gamma$\, coupling \cite{Fabbrichesi:2017vma}.

In this paper, we would like to show that the hyperon sector potentially offers a competitive
window to access \,$s\to d\overline\gamma$.\,
Generally the quark transition \,$s\to d\slashed E$\, gives rise to the FCNC decays of hyperons
into a lighter baryon plus missing energy ($\slashed E$) which is carried away by one or more
invisible particles.
If these are a pair of spin-1/2 fermions or spinless bosons, it is presently possible for NP
to enhance the branching fractions of the hyperon modes considerably above the SM expectations
to levels discoverable by ongoing or future searches~\cite{Tandean:2019tkm,Li:2019cbk}.
Particularly, new results on these decays may be forthcoming from the BESIII
experiment~\cite{Li:2016tlt,Ablikim:2019hff}, supplying information on potential NP in
\,$s\to d\slashed E$\, complementary to that gained from quests for their kaon counterparts.
If the missing energy is carried away by a massless dark photon instead, the final state has only
two particles, and consequently the prospects for BESIII pursuing these two-body modes are
expected to be comparatively better as long as their branching fractions are not too small.
As we will see shortly, they can indeed be sizable in some scenarios recently proposed in
the literature, implying that the two-body decays may be testable with the existing hyperon
data and upcoming measurements.

In the main processes under consideration a massless dark photon, $\overline\gamma$, is emitted
on-shell due to its dipole-type flavor-changing couplings to the $d$ and $s$ quarks brought
about by loop diagrams involving heavy nonstandard particles.
This is described by the dimension-five operators in the effective Lagrangian
\begin{align} \label{Ldsg}
{\cal L}_{ds\bar\gamma}^{} & \,=\, -\overline d\sigma^{\mu\nu}
(\mbox{\small$\mathbb C$}+\gamma_5^{}\mbox{\small$\mathbb C$}_5^{})s \bar F_{\mu\nu}^{}
\,+\, {\rm H.c.} \,,
\end{align}
where $\mbox{\small$\mathbb C$}$ and $\mbox{\small$\mathbb C$}_5$ are constants having
the dimension of inverse mass and dependent on the details of the underlying NP model,
\,$\sigma^{\mu\nu}=(i/2)[\gamma^\mu,\gamma^\nu]$,\, and
\,$\bar F_{\mu\nu}^{}=\partial_\mu\bar A_\nu-\partial_\nu\bar A_\mu$\, denotes
the field strength tensor of $\overline\gamma$.

Our hyperon decays of interest are \,$\mathfrak B\to\mathfrak B'\overline\gamma$\, with
\,$\mathfrak{BB}'=\Lambda n,\Sigma^+p,\Xi^0\Lambda,\Xi^0\Sigma^0,\Xi^-\Sigma^-$, all of the baryons
having spin 1/2, and \,$\Omega^-\to\Xi^-\overline\gamma$,\, where $\Omega^-$ has spin 3/2.
For these modes, to calculate the baryonic parts of the amplitudes induced by
${\cal L}_{ds\bar\gamma}$, we need to know the matrix elements
\,$\langle{\mathfrak B}'|\overline d\sigma^{\mu\nu}(1,\gamma_5^{})s|{\mathfrak B}\rangle$\, and
\,$\langle\Xi^-|\overline d\sigma^{\mu\nu}(1,\gamma_5^{})s|\Omega^-\rangle$\, contracted with
the dark photon's momentum $\overline q$ and polarization $\overline\varepsilon$, which satisfy
the gauge condition \,$\overline\varepsilon\cdot\overline q=0$.\,
We additionally have \,$\bar q^2=0$\, due to $\overline\gamma$ being on-shell and
\,$\overline q=p_{\mathfrak B}^{}-p_{\mathfrak B'}^{}=p_\Omega^{}-p_\Xi^{}$,\,
the momentum difference between the initial and final baryons.
It follows that we can express
\begin{align} \label{<B'B>}
\langle{\mathfrak B}'|\overline di\sigma^{\mu\nu}(\mbox{\small$\mathbb C$} + \gamma_5^{}
\mbox{\small$\mathbb C$}_5^{})s|{\mathfrak B}\rangle\overline\varepsilon_\mu^*\overline q_\nu^{}
& \,=\, \frac{f_{\mathfrak B'\mathfrak B}^{}}{2}\, \bar u_{\mathfrak B'}^{} \big(
\!\!\not\!\overline\varepsilon^* \slashed{\overline q} - \slashed{\overline q}\!\not\!
\overline\varepsilon^* \big) (\mbox{\small$\mathbb C$}+\gamma_5^{}\mbox{\small$\mathbb C$}_5^{})
u_{\mathfrak B}^{} \,,
\\ \label{<XO>}
\langle\Xi^-|\overline di\sigma^{\mu\nu}(\mbox{\small$\mathbb C$} + \gamma_5^{}
\mbox{\small$\mathbb C$}_5^{})s|\Omega^-\rangle\overline\varepsilon_\mu^*\overline q_\nu^{}
& \,=\, \bar u_\Xi^{} \Bigg( \frac{f_{\Xi\Omega}^{}}{2} + \frac{\tilde f_{\Xi\Omega}^{}\,
\slashed p{}_\Omega^{}}{2m_{\Omega^-}^{}} \Bigg) \big( \!\!\not\!\overline\varepsilon^*
\overline q_\tau^{} - \slashed{\overline q}\, \overline\varepsilon_\tau^* \big)
(\gamma_5^{}\mbox{\small$\mathbb C$} + \mbox{\small$\mathbb C$}_5^{}) u_\Omega^\tau \,, ~~~
\end{align}
where $f_{\mathfrak B'\mathfrak B}^{}$, $f_{\Xi\Omega}^{}$, and $\tilde f_{\Xi\Omega}^{}$ represent
form factors evaluated at \,$\overline q^2=0$\, and $u_X^{}$ $\big(u_\Omega^\mu\big)$ stands for
the Dirac (Rarita-Schwinger) spinor of $X$ $(\Omega^-)$.
With these formulas, we derive the decay rates, arriving at
\begin{align} \label{GB2Bg'}
\Gamma_{{\mathfrak B}\to{\mathfrak B}'\overline\gamma}^{} & \,=\,
\frac{\Delta_{\mathfrak{BB}'}^6\, f_{\mathfrak B'\mathfrak B}^2}{2\pi m_{\mathfrak B}^3}
\big(|\mbox{\small$\mathbb C$}|^2+|\mbox{\small$\mathbb C$}_5|^2\big) \,,
\nonumber \\
\Gamma_{\Omega^-\to\Xi^-\overline\gamma}^{} & \,=\, \frac{4 \Delta_{\Omega^-\Xi^-}^6
\big(f_{\Xi\Omega}^{} m_{\Omega^-}^{}+\tilde f_{\Xi\Omega}^{}m_{\Xi^-}^{}\big)\raisebox{1pt}{$^2$}
+ \Delta_{\Omega^-\Xi^-}^8 \big(\tilde f_{\Xi\Omega}^2-f_{\Xi\Omega}^2\big)}{96\pi m_{\Omega^-}^5}
\big(|\mbox{\small$\mathbb C$}|^2+|\mbox{\small$\mathbb C$}_5|^2\big) \,,
\end{align}
where \,$\Delta_{XY}=\sqrt{m_X^2-m_Y^2}$.\,
As regards $f_{\mathfrak B'\mathfrak B}^{}$, $f_{\Xi\Omega}^{}$, and $\tilde f_{\Xi\Omega}^{}$,
it turns out that matrix elements analogous to those in Eqs.\,\,(\ref{<B'B>}) and (\ref{<XO>})
were estimated a while ago in an investigation~\cite{Gilman:1978bg} on the contributions of
the transition \,$s\to d\gamma$\, to hyperon radiative weak decays,
\,${\mathfrak B}\to{\mathfrak B}'\gamma$\, \mbox{and \,$\Omega^-\to\Xi^-\gamma$,\,}
involving the ordinary photon.\footnote{It is worth noting that (\ref{<B'B>}) and (\ref{<XO>})
can be shown to be consistent in form with the general gauge-invariant amplitudes known in
the literature for these hyperon radiative weak
decays~\cite{Behrends:1958zz,Eeg:1983zi,Safadi:1987qq,Eilam:1996vs}.}
With appropriate changes, we can apply the results found therein~\cite{Gilman:1978bg} to
the corresponding processes with a massless dark photon, thus obtaining
\begin{align} \label{ff}
f_{\mathfrak B'\mathfrak B}^{} & \,=\, {\cal C}_{\mathfrak B'\mathfrak B}^{}
\sqrt{\frac{m_{\mathfrak B'}^{}}{m_{\mathfrak B}^{}}} \,, &
f_{\Xi\Omega}^{} & \,=\, \tilde f_{\Xi\Omega}^{} \,=\,
\frac{4\sqrt{m_\Omega^{}m_\Xi^{}}}{m_\Omega^{}+m_\Xi^{}} \,. ~~~~~
\end{align}
For all of these modes, \,${\mathfrak B}_{\rm i}^{}\to{\mathfrak B}_{\rm f}^{}\overline\gamma$,\,
combining Eq.\,(\ref{ff}) with Eq.\,(\ref{GB2Bg'}) we can write the rates as
\begin{align} \label{GBi2Bfg'}
\Gamma_{{\mathfrak B}_{\rm i}^{}\to{\mathfrak B}_{\rm f^{}}\overline\gamma}^{} & \,=\,
\frac{\Delta_{{\mathfrak B}_{\rm i}{\mathfrak B}_{\rm f}}^6\,
{\cal C}_{{\mathfrak B}_{\rm f}{\mathfrak B}_{\rm i}}^2\, m_{{\mathfrak B}_{\rm f}}^{}}
{2\pi\, m_{{\mathfrak B}_{\rm i}}^4}
\big(|\mbox{\small$\mathbb C$}|^2+|\mbox{\small$\mathbb C$}_5|^2\big) \,, ~~~
\end{align}
and in Table\,\,\ref{CBB'} we have listed the values of
${\cal C}_{{\mathfrak B}_{\rm f}{\mathfrak B}_{\rm i}}^2$, which were determined in
Ref.\,\cite{Gilman:1978bg} with quark-model SU(6) wave functions and for which
we have assumed overlap factors of~1.

Since $\mbox{\small$\mathbb C$}$ and $\mbox{\small$\mathbb C$}_5$ are generally independent
of each other, for simplicity hereafter we assume that $\mbox{\small$\mathbb C$}_5$ is absent,
which is also the case in the particular examples we look at below.
For each of the hyperon modes, we can then evaluate the corresponding branching fraction
${\cal B}\big({\mathfrak B}_{\rm i}^{}\to{\mathfrak B}_{\rm f}^{}\overline\gamma\big)$
once $\mbox{\small$\mathbb C$}$ is numerically given.
For later convenient use, we have collected in Table\,\,\ref{BB2B'g'} the numbers for
${\cal B}\big({\mathfrak B}_{\rm i}^{}\to{\mathfrak B}_{\rm f}^{}\overline\gamma\big)$
divided by $|\mbox{\small$\mathbb C$}|^2$, after employing the measured masses from
Ref.\,\cite{Tanabashi:2018oca}.
In light of the successful quark-model predictions of other baryonic quantities which rely on
the same assumption of a single-quark operator transforming like a component of the quark
spin~\cite{Gilman:1978bg}, we expect that the estimates displayed in this table are
good to within factors of 2.

\begin{table}[!b] \smallskip
\begin{tabular}{|c||c|c|c|c|c|c|} \hline
${\mathfrak B}_{\rm f}^{}{\mathfrak B}_{\rm i}^{}\vphantom{\int_|^\int}$ & ~$n\Lambda$~ &
~$p\Sigma^+$~ & ~$\Lambda\Xi^0$~ & ~$\Sigma^0\Xi^0$~ & ~$\Sigma^-\Xi^-$~ & ~$\Xi^-\Omega^-$~
\\ \hline\hline
~${\cal C}_{{\mathfrak B}_{\rm f}{\mathfrak B}_{\rm i}}^2$~ & \large~$\frac{3}{2}$~ &
\large$\frac{1}{9}$ & \large$\frac{1}{6}$ & \large$\frac{25}{18}$ & \large$\frac{25}{9}$ &
\large$\frac{4}{3}_{\vphantom{\int_|}}^{\vphantom{\int^|}}$ \\ \hline
\end{tabular}
\caption{Values of ${\cal C}_{{\mathfrak B}_{\rm f}{\mathfrak B}_{\rm i}}^2$ in
Eq.\,(\ref{GBi2Bfg'}) for \,${\mathfrak B}_{\rm f}^{}{\mathfrak B}_{\rm i}^{}=n\Lambda,p\Sigma^+,
\Lambda\Xi^0,\Sigma^0\Xi^0,\Sigma^-\Xi^-,\Xi^-\Omega^-$\,
from Ref.\,\cite{Gilman:1978bg}.} \label{CBB'}
\end{table}
\begin{table}[!b] \smallskip
\begin{tabular}{|c||c|c|c|c|c|c|} \hline
~Decay mode~ & $\Lambda\to n\overline\gamma\vphantom{\int_|^\int}$ &
$\Sigma^+\to p\overline\gamma$ & $\Xi^0\to\Lambda\overline\gamma$ &
$\Xi^0\to\Sigma^0\overline\gamma$ & $\Xi^-\to\Sigma^-\overline\gamma$ &
$\Omega^-\to\Xi^-\overline\gamma$
\\ \hline \hline
$\displaystyle\frac{\cal B}{|\mbox{\small$\mathbb C$}|^2}
_{\vphantom{\int_|}}^{\vphantom{\int^|}}$\,({\footnotesize$\rm GeV^2$}) &
~$2.75\times10^{12}$~ & ~$1.54\times10^{11}$~ & ~$4.95\times10^{11}$~ &
~$1.12\times10^{12}$~ & ~$1.32\times10^{12}$~ & ~$5.18\times10^{12}$~
\\ \hline
\end{tabular}
\caption{The branching fractions $\cal B$ of
\,${\mathfrak B}_{\rm i}^{}\to{\mathfrak B}_{\rm f}^{}\overline\gamma$\, divided by
$|\mbox{\small$\mathbb C$}|^2$ in the \,$\mbox{\small$\mathbb C$}_5=0$\, case.\label{BB2B'g'}}
\end{table}

To see how large
${\cal B}\big({\mathfrak B}_{\rm i}^{}\to{\mathfrak B}_{\rm f}^{}\overline\gamma\big)$
may be at present without restrictions from hyperon data, we now consider, as benchmarks,
different maximal values of $|\mbox{\small$\mathbb C$}|$ in two scenarios recently discussed
in the literature.
Although they are variations of the same simplified model of NP, the two different sets of
predictions which they make will serve to illustrate how hyperon and kaon measurements
together could help distinguish various NP possibilities.
Therefore, similar analyses could be performed for other models.
In both of these scenarios, proposed in Refs.\,\,\cite{Fabbrichesi:2017vma,Fabbrichesi:2019bmo},
the $ds\overline\gamma$ interactions originate from loop diagrams involving new particles
comprising massive fermions which are SM singlets as well as heavy scalar bosons which are
triplets under color SU(3) and some of which are doublets under the SM SU(2)$_L$.
The new fermions and bosons are all charged under U(1)$_D$ and have Yukawa-like interactions
with the $d$ and $s$ quarks, which allows the dimension-five operator for
\,$s\to d\overline\gamma$\, to be constructed.
Since our aim here is to examine the implications of this coupling for the hyperon modes, once
it has been subject to constraints from other sectors, we will not dwell further on the details
of the underlying NP model.
Rather, we will simply take most of the relevant results provided in
Refs.\,\,\cite{Fabbrichesi:2017vma,Fabbrichesi:2019bmo} at face value and employ them to compute
the hyperon decay rates.

In the first scenario the $ds\overline\gamma$ coupling constant is
given by~\cite{Fabbrichesi:2017vma}
\begin{align} \label{C}
\mbox{\small$\mathbb C$} & \,=\, \frac{e_D^{}\xi}{64\pi^2\tilde\Lambda} \,,
\end{align}
where $e_D^{}$ parametrizes the dark photon's interaction strength, $\xi$ is a product of two
common Yukawa couplings between the new particles and SM quarks, and $\tilde\Lambda$ is
the effective heavy mass scale of the dark sector.
The Yukawa interactions also bring about box-diagram contributions to the kaon-mixing
quantity $\Delta m_K^{}$.
Their impact can be numerically expressed in terms of the ratio $\xi/\tilde\Lambda$
as~\cite{Fabbrichesi:2017vma}
\begin{align} \label{dmk}
\Delta m_K^{\textsc{np}} & \,=\, 8.47\times10^{-13}{\rm\,TeV}^3\, \frac{\xi^2}{\tilde\Lambda^2} \,,
\end{align}
which is dominated by the effects of four-quark operators that have chirally enhanced
matrix elements between the $K^0$ and $\bar K^0$ states.
To constrain $\xi/\tilde\Lambda$, the new contribution in Eq.\,(\ref{dmk}) is
required~\cite{Fabbrichesi:2017vma} to be less than 30\% of its experimental counterpart,
namely \,$\Delta m_K^{\textsc{np}}<0.3\Delta m_K^{\rm exp}$,\, where
\,$\Delta m_K^{\rm exp}=3.484\times10^{-15}$\,GeV \cite{Tanabashi:2018oca}.
This constitutes the main restraint on $\xi/\tilde\Lambda$.
With the size of $e_D^{}$ following from the choice \,$\alpha_D^{}=e_D^2/(4\pi)=0.1$
\cite{Fabbrichesi:2017vma}, putting things together we then obtain from Eq.\,(\ref{C})
\begin{align} \label{I}
|\mbox{\small$\mathbb C$}| \,<\, 2.0\times10^{-9}\rm~GeV^{-1} \,.
\end{align}
Combining this with the entries in Table\,\,\ref{BB2B'g'} yields the maximal branching
fractions ${\cal B}_{\rm max}$ of the hyperon modes shown in the second row of
Table\,\,\ref{numBB2B'g'} and labeled [I].

\begin{table}[t]
\begin{tabular}{|c||c|c|c|c|c|c|} \hline
~Decay mode~ & $\Lambda\to n\overline\gamma\vphantom{\int_|^\int}$ &
$\Sigma^+\to p\overline\gamma$ & $\Xi^0\to\Lambda\overline\gamma$ &
$\Xi^0\to\Sigma^0\overline\gamma$ & ~$\Xi^-\to\Sigma^-\overline\gamma$~ &
~$\Omega^-\to\Xi^-\overline\gamma$~
\\ \hline \hline
$\cal B_{\rm max}{}_{\vphantom{\int}}^{\vphantom{\int^0}}$ [I] & ~$1.1\times10^{-5}$~ &
~$6.0\times10^{-7}$~ & ~$1.9\times10^{-6}$~ & ~$4.3\times10^{-6}$~ &
\,$5.1\times10^{-6}$\, & \,$2.0\times10^{-5}$\,
\\ \hline
$\cal B_{\rm max}{}_{\vphantom{\int}}^{\vphantom{\int^0}}$ [II] & \,$6.7\times10^{-4}$\, &
\,$3.8\times10^{-5}$\, & \,$1.2\times10^{-4}$\, & \,$2.7\times10^{-4}$\, &
\,$3.3\times10^{-4}$\, & \,$1.3\times10^{-3}$\,
\\ \hline
\end{tabular}
\caption{\label{numBB2B'g'}The maximal branching fractions of
\,${\mathfrak B}_{\rm i}^{}\to{\mathfrak B}_{\rm f}^{}\overline\gamma$\, with
the $|\mbox{\small$\mathbb C$}|$ values from [I] Eq.\,(\ref{I}) and [II]~Eq.\,(\ref{II}).}
\end{table}

In the second scenario the $ds\overline\gamma$ coupling constant has
the form \cite{Fabbrichesi:2019bmo}
\begin{align} \label{ii}
\mbox{\small$\mathbb C$} & \,=\, \frac{e_D^{}{\cal D}_M^{}}{2\tilde\Lambda} \,,
\end{align}
where ${\cal D}_M^{}$ contains a product of a couple of Yukawa couplings between the new
particles and the $d$ and $s$ quarks and is related to the kaon-mixing parameter by
\,$\Delta m_K^{\textsc{np}}=32\pi^2 {\cal D}_M^2 f_K^2 m_{K^0}/(3\tilde\Lambda^2)$,\,
with \,$f_K^{}=159.8$ MeV\, being the kaon decay constant and
\,$m_{K^0}=497.6$ MeV \cite{Tanabashi:2018oca}.
In~this instance, unlike the preceding one, the $K^0$-$\bar K^0$ matrix elements of
the contributing four-quark operators are not chirally enhanced.
To be consistent with the previous example, we \mbox{demand
\,$\Delta m_K^{\textsc{np}}<0.3\Delta m_K^{\rm exp}$},\, leading to
\begin{align}
\frac{{\cal D}_M^2}{\tilde\Lambda^2} & \,<\, 7.8\times10^{-16}\rm~GeV^{-2} \,,
\end{align}
which is a more stringent limit than that found in Ref.\,\cite{Fabbrichesi:2019bmo} from
the condition \,$\Delta m_K^{\textsc{np}}<\Delta m_K^{\rm exp}$.\,
Applying this to Eq.\,(\ref{ii}), with \,$\alpha_D^{}=0.1$\, as before, we then get
\begin{align} \label{II}
|\mbox{\small$\mathbb C$}| \,<\, 1.6\times10^{-8}\rm~GeV^{-1} \,.
\end{align}
This translates into the maximal branching fractions ${\cal B}_{\rm max}$ of the hyperon
modes listed in the last row of Table\,\,\ref{numBB2B'g'} and labeled [II].

As indicated above, the differences between the two sets of predictions in
Table\,\,\ref{numBB2B'g'} reflect the fact that the details of the underlying NP in
the two cases affect kaon mixing differently.
It follows that quests for the hyperon modes could come up with information pertinent
for testing NP models.
The future measurement on \,$K^-\to\pi^-\pi^0\overline\gamma$,\, which arises from the same
\,$s\to d\overline\gamma$\, transition, would likely be important as well according to
Ref.\,\cite{Fabbrichesi:2017vma}.

As pointed out earlier, the hyperon decays may be searched for in the BESIII
experiment~\cite{Li:2016tlt}, which has produced copious $\Lambda$, $\Sigma$, $\Xi$, and
$\Omega$ hyperons \cite{Ablikim:2019hff}.
For their FCNC decays with missing energy carried away by a pair of invisible
particles, the proposed BESIII sensitivity levels for the branching fractions of
\,$\Lambda\to n\slashed E$,\, $\Sigma^+\to p\slashed E$,\, $\Xi^0\to\Lambda\slashed E$,\,
$\Xi^0\to\Sigma^0\slashed E$,\, and \,$\Omega^-\to\Xi^-\slashed E$\,
\mbox{are \,$3\times10^{-7}$},\, $4\times10^{-7}$,\, $8\times10^{-7}$,\, $9\times10^{-7}$,\, and
\,$2.6\times10^{-5}$,\, respectively~\cite{Li:2016tlt}.
Since our decays of concern,
\,${\mathfrak B}_{\rm i}^{}\to{\mathfrak B}_{\rm f}^{}\overline\gamma$,\, are two-body ones,
BESIII would have greater sensitivity to them, being able to detect them more efficiently than
the three-body ones.
This is especially consequential for NP models which generate predictions comparable
to those displayed in Table\,\,\ref{numBB2B'g'}.

As we await possibly upcoming outcomes of the direct searches for
\,${\mathfrak B}_{\rm i}^{}\to{\mathfrak B}_{\rm f}^{}\overline\gamma$\, from BESIII and other
(future) experiments, it turns out that we can already extract approximate bounds on their
branching fractions indirectly from the data on the observed channels quoted by the Particle
Data Group~\cite{Tanabashi:2018oca}.
To do so, for each of these decaying hyperons, we subtract from unity the sum of the PDG
branching-fraction numbers with their errors (at 2 sigmas) combined in quadrature.
Thus, for the yet-unobserved decay modes of $\Lambda$, $\Sigma^+$, $\Xi^0$, $\Xi^-$, and
$\Omega^-$, we arrive at the upper limits of \,$1.4\%$,\, $8.0\times10^{-3}$,\,
$3.4\times10^{-4}$,\, $8.3\times10^{-4}$,\, and \,$1.6\%$, respectively.\,
Evidently, they accommodate most of the corresponding predictions in Table\,\,\ref{numBB2B'g'},
the exceptions being the two $\Xi^0$ results together in the bottom row.
This suggests that the second scenario considered above may already be in tension with the present
hyperon data if its $ds\overline\gamma$ coupling constant is not significantly lower than
the bound in Eq.\,(\ref{II}).
Had we imposed the weaker \mbox{condition \,$\Delta m_K^{\textsc{np}}<\Delta m_K^{\rm exp}$},\,
the resulting upper values of $|\mbox{\small$\mathbb C$}|$ in this scenario would be disfavored
more strongly.

In conclusion, we have entertained the possibility that massless dark photons exist and
have nonnegligible flavor-changing dipole-type couplings to down-type light quarks.
These kind of interactions induce FCNC hyperon decays into a lighter baryon plus missing energy
carried away by the massless dark photon.
We demonstrate that these hyperon modes can be a powerful tool to probe the underlying
$s\to d\overline\gamma$ transition, perhaps more so than what kaon decays could offer.
We illustrate this with a couple of examples from recent studies in the literature which
take into account the current restrictions from the kaon sector.
Our findings reveal that one of these two scenarios might already be in tension with the indirect
bounds on hyperon decays with missing energy inferred from the available hyperon data.
Further tests on the hyperon modes will expectedly come from ongoing experiments such as
BESIII and future facilities such as super charm-tau factories.
In the near future, BESIII will likely be in a position to discover one or more of these
hyperon modes or, if not, set stringent limitations on the $ds\overline\gamma$ interactions.

\acknowledgements

We would like to thank Hai-Bo Li for information on experimental matters.
This research was supported in part by the MOST (Grant No. MOST 106-2112-M-002-003-MY3).

\end{document}